\documentclass[preprint,showpacs,showkeys,preprintnumbers,amsmath,amssymb,superscriptaddress]{revtex4}
\usepackage{graphicx}
\usepackage{dcolumn}
\usepackage{bm}

\begin{document}
\title{Hawking radiation as tunneling from a Vaidya black hole in noncommutative gravity}

\author{S. Hamid Mehdipour}

\email{mehdipour@iau-lahijan.ac.ir}

\affiliation{Islamic Azad University, Lahijan Branch, P. O. Box
1616, Lahijan, Iran}

\date{\today}

\begin{abstract}
In the context of a noncommutative model of coordinate coherent
states, we present a Schwarzschild-like metric for a Vaidya solution
instead of the standard Eddington-Finkelstein metric. This leads to
the appearance of an exact $(t - r)$ dependent case of the metric.
We analyze the resulting metric in three possible causal structures.
In this setup, we find a zero remnant mass in the long-time limit,
i.e. an instable black hole remnant. We also study the tunneling
process across the quantum horizon of such a Vaidya black hole. The
tunneling probability including the time-dependent part is obtained
by using the tunneling method proposed by Parikh and Wilczek in
terms of the noncommutative parameter $\sigma$. After that, we
calculate the entropy associated to this noncommutative black hole
solution. However the corrections are fundamentally trifling; one
could respect this as a consequence of quantum inspection at the
level of semiclassical quantum gravity.
\end{abstract}


\maketitle

\section{\label{sec:1}Introduction}
There have been many paradigms for the noncommutative field theory
based on the Weyl-Wigner-Moyal $\ast$-product \cite{wey} which fail
to find a way for solving the subsequent problems, such as Lorentz
invariance breaking, nonunitarity and UV divergences of quantum
field theory. Recently, Smailagic and Spallucci \cite{sma} suggested
a noncommutative model of coordinate coherent states (CCS) which
could be released from the above problems. Their findings were
acquired by beginning with an innovative method to noncommutative
geometry after a long period of time. Using the CCS approach, the
authors in \cite{nic} derived exact solutions of the Einstein
equations for a static, spherically symmetric, asymptotically flat,
minimal width, mass/energy distribution localized near the origin;
as a result there is no curvature singularity at the origin. In this
model, the pointlike structure of mass $M$, instead of being
completely localized at a point, is portrayed by a smeared structure
throughout a region of linear size $\sqrt{\sigma}$. The
characteristic energy or inverse length scale related to the
noncommutativity effects possibly and most rationally would have a
natural value of order of the Planck scale. In fact, most of the
phenomenological studies of the noncommutativity models expect that
the noncommutative energy scale cannot lie far above the TeV scale
\cite{hin}. Since the fundamental Planck scale in models with large
extra dimensions becomes as small as a TeV in order to solve the
hierarchy problem \cite{ant}, therefore, depending on the models, it
is feasible to set the noncommutativity effects in a $1-10$ TeV
regime, etc. \cite{riz}.

A radiation spectrum of an evaporating black hole, which is closely
comparable to the blackbody radiation spectrum, can be illustrated
by a characteristic temperature known as the Hawking temperature
\cite{haw}. Hawking's method unfortunately yields a nonunitarity of
quantum theory, which maps a pure state to a mixed state, due to the
purely thermal essence of the spectrum. In 2000, Parikh and Wilczek
\cite{par1} presented a new approach on the basis of null geodesics
to draw out the Hawking radiation via tunneling through the quantum
horizon. In this approach, the form of the black hole radiation
spectrum is modified as a result of incorporation of backreaction
effects. From another point of view, Shankaranarayanan {\it et al}.
performed the tunneling process to get the Hawking temperature in
different coordinates within a complex paths approach \cite{sha}.
The tunneling procedure illuminates the fact that the modified
radiation spectrum is not accurately thermal and this yields the
unitarity of underlying quantum theory \cite{par2}.

In this paper, we would like to suggest a new formulation of
noncommutativity of coordinates for a Vaidya black hole which is
performed by a Gaussian distribution of coherent states. This type
of black hole is considered as an illustration of a more practical
case because it is a time-dependent lessening mass caused by the
evaporation process. This trend continues to proceed the
Parikh-Wilczek tunneling procedure through the event horizon of such
a noncommutative-inspired Vaidya black hole.

The organization of this paper is as follows. In Sec.~\ref{sec:2},
the influence of noncommutativity in the framework of coordinate
coherent states for a Vaidya metric is investigated. In this manner,
an exact $(t - r)$ dependence solution is obtained. In
Sec.~\ref{sec:3}, we study the Parikh-Wilczek tunneling for such a
Vaidya solution. The tunneling amplitude at which massless particles
tunnel across the event horizon is computed. Finally, the
conclusions of the work in this paper are
summarized in Sec.~\ref{sec:4}.\\

\section{\label{sec:2}Disappearance of Black Hole Remnant}
As mentioned briefly in the Introduction, the simple idea of a
pointlike particle becomes physically irrelevant and should be
replaced with a minimal width Gaussian distribution of mass/energy,
corresponding to the principles of quantum mechanics \cite{nic}. The
program we choose here is to perform an analysis which provides a
solution in the case of a nonstatic, spherically symmetric,
asymptotically flat, minimal width, Gaussian distribution of
mass/energy whose noncommutative size is characterized by the
parameter $\sqrt{\sigma}$. For this purpose, the mass/energy
distribution should be displayed by a smeared delta function
\begin{equation}
\label{mat:1}\rho_{\sigma}=\frac{M}
{(4\pi\sigma)^{\frac{3}{2}}}e^{-\frac{r^2}{4\sigma}},
\end{equation}
where, in this approach, $\rho_{\sigma}=\rho_{\sigma}(t,r)$ and
$M=M(t,r)$ are functions of both time and radius. The dynamics for
the black hole mass with evaporation is a persistent problem. In the
study of black hole evaporation, there is a significant point in
which the black hole mass reduces as a backreaction of the Hawking
radiation. Since a nonstatic and spherically symmetric spacetime
depends on an arbitrary dynamical mass function, it can be properly
demonstrated by a Vaidya solution \cite{vai} that seems to be the
favored option. In this work, due to deriving an exact $(t-r)$
dependent case of the metric, we study the Schwarzschild-like metric
for the Vaidya solution instead of a standard Eddington-Finkelstein
metric.

In this paper we want to generalize the Vaidya metric derived by
Farley and D'Eath \cite{far} to the noncommutative model of CCS. The
general spherically symmetric Vaidya spacetime in $\{x^\mu\}=\{t, r,
\theta, \phi\}$ coordinates, ($\mu=0,1,2,3$), in the presence of
$(t-r)$ dependent mass sources can be written as \cite{far}
\begin{equation}
\label{mat:2}ds^2=-e^{2\Psi(t,r)}F(t,r)dt^2+F^{-1}(t,r)dr^2+r^2d\Omega^2,
\end{equation}
where $d\Omega^2 = d\theta^2 + sin^2\theta \,d\phi^2$, and
$e^{2\Psi(t,r)}=\left(\frac{\dot{M}}{\chi(M)}\right)^2$. Here
$\chi(M)$ is the arbitrary positive function of $t$ and $r$. In the
following, due to the mathematical intricacy of Einstein field
equations and in order to make the problem obedient, we frequently
impose the particular cases, e.g. $\chi(M) = -\dot{M}$
($\dot{M}<0$), the overdot abbreviates $\frac{\partial}{\partial
t}$. This metric looks like the Schwarzschild spacetime, except that
the role of the Schwarzschild mass is taken by a mass function $M(t,
r)$, which changes exceedingly gradually concerning both $t$ and $r$
in the spacetime region containing the outgoing radiation
\cite{ste}. The corresponding geometry in this region including the
radially outgoing radiation is, therefore, the slowly varying Vaidya
type.

In order to determine the mass function, we consider the covariant
conservation condition $T^{\mu\nu}_{\quad;\nu}=0$, which yields the
explicit result
\begin{equation}
\label{mat:3}\partial_tT_{t}^{\,\,\,\,t}+\partial_rT_{r}^{\,\,\,\,r}+\frac{1}{2}g^{tt}\partial_rg_{tt}(T_{r}^{\,\,\,\,r}-T_{t}^{\,\,\,\,t})+
\frac{1}{2}g^{rr}\partial_tg_{rr}(T_{r}^{\,\,\,\,r}-T_{t}^{\,\,\,\,t})+g^{\theta\theta}\partial_rg_{\theta\theta}(T_{r}^{\,\,\,\,r}-
T_{\theta}^{\,\,\,\,\theta})=0.
\end{equation}
The Schwarzschild-like condition $g_{tt}=-g_{rr}^{-1}$ will require
that $T_{t}^{\,\,\,\,t}=T_{r}^{\,\,\,\,r}=-\rho_{\sigma}$, and then
the above relation leads to a solution for
$T_{\theta}^{\,\,\,\,\theta}$ which reads \footnote{We set the
fundamental constants equal to unity; $ \hbar= c = G = 1$.}
\begin{equation}
\label{mat:4}T_{\theta}^{\,\,\,\,\theta}=\rho_\sigma\left(\frac{r^2}{4\sigma}-\frac{r}{2M}\Big(\dot{M}+M'\Big)-1\right),
\end{equation}
where the prime abbreviates $\frac{\partial}{\partial r}$. The
nonzero components of the Einstein field equations $G_{\mu\nu}=8\pi
T_{\mu\nu}$ give the following equations:
\begin{equation}
\label{mat:5}F'r+F+8\pi r^2\rho_\sigma-1=0,
\end{equation}
\begin{equation}
\label{mat:6}\dot{F}+8\pi rF^2T_{t}^{\,\,\,\,r}=0,
\end{equation}
\begin{equation}
\label{mat:7}rF''F^3-2r\dot{F}^2+r\ddot{F}F+2F^3F'-16\pi
rF^3T_{\theta}^{\,\,\,\,\theta}=0,
\end{equation}
\begin{equation}
\label{mat:8}T_{t}^{\,\,\,\,r}=T_{r}^{\,\,\,\,t}, \quad
\textmd{and}\quad
T_{\phi}^{\,\,\,\,\phi}=T_{\theta}^{\,\,\,\,\theta}.
\end{equation}
Now, one can describe the self-gravitating, anisotropic matter
source through a fluid-type $T_{\mu}^{\,\,\,\,\nu}$ of the following
form:
\begin{equation}\label{mat:9} T_{\mu}^{\,\,\,\,\nu} = \left(
\begin{array}{cccc}
T_{t}^{\,\,\,\,t} & T_{t}^{\,\,\,\,r} & 0  & 0\\
 T_{r}^{\,\,\,\,t}& T_{r}^{\,\,\,\,r} & 0 & 0\\
 0 & 0 & T_{\theta}^{\,\,\,\,\theta} & 0 \\
 0 & 0 & 0 & T_{\phi}^{\,\,\,\,\phi}
\end{array} \right).
\end{equation}
This type of energy-momentum tensor is slightly atypical due to the
fact that $T_{\mu}^{\,\,\,\,\nu}$ deviates from the conventional
perfect fluid form including the isotropic pressure terms. However,
according to the slowly varying Vaidya form ($\dot{M}\ll 1$ and
$M'\ll 1$) it is easy to show that the pressure terms
$T_{r}^{\,\,\,\,r}$ and $T_{\theta}^{\,\,\,\,\theta}$ are different
only surrounded by a few $\sqrt{\sigma}$ from the origin and the
perfect fluid condition is recovered for larger distances.

\begin{figure}[htp]
\begin{center}
\includegraphics{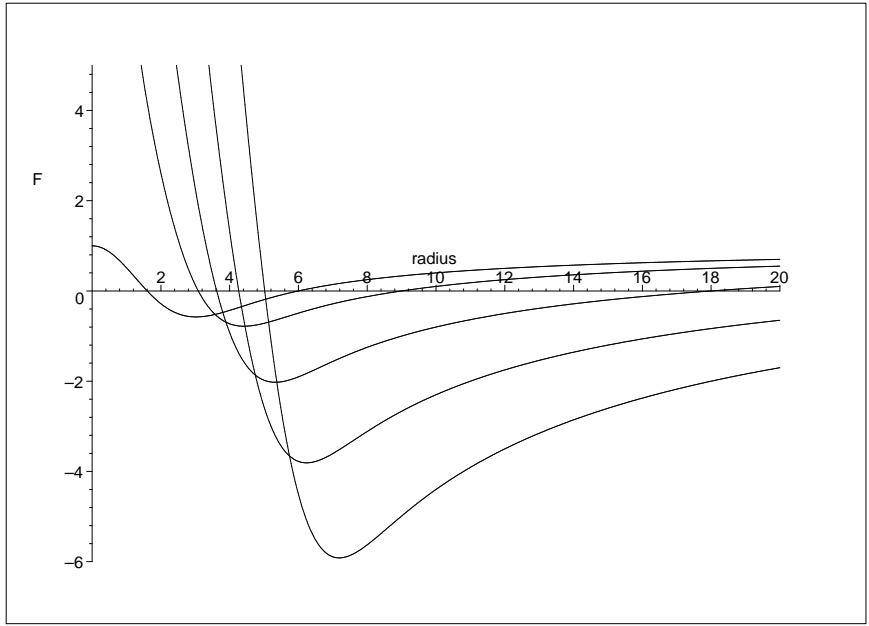}
\end{center}
\vspace{8 cm}
 \caption{\scriptsize {The temporal component of the
metric versus $\frac{r}{\sqrt{\sigma}}$ for different values of
$\frac{t}{\sqrt{\sigma}}$ with a sufficiently large amount of
initial mass ($M_I>M_0$), e.g. $M_I=3.00\sqrt{\sigma}$. On the
right-hand side of the figure, curves are marked from top to bottom
by $t = 0,~ 1.00\sqrt{\sigma},~ 2.00\sqrt{\sigma},~
3.00\sqrt{\sigma},$ and $4.00\sqrt{\sigma}$. This figure shows that,
in the long-time limit, the distance between the horizons is
increased.}}
 \label{fig:1}
\end{figure}

\begin{figure}[htp]
\begin{center}
\includegraphics{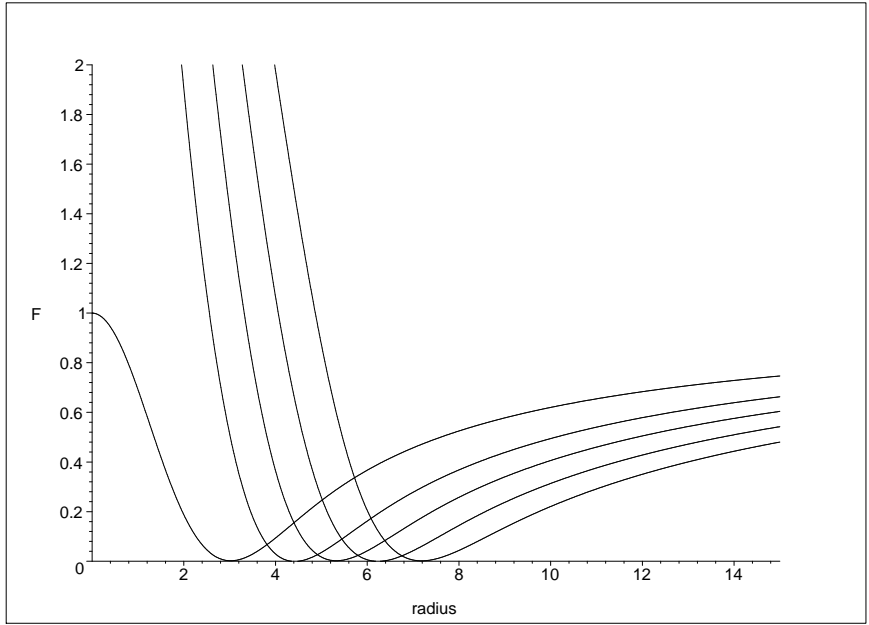}
\end{center}
\vspace{8 cm}
 \caption{\scriptsize {The temporal component of the
metric versus $\frac{r}{\sqrt{\sigma}}$ for different values of
$\frac{t}{\sqrt{\sigma}}$ under the condition that $M_I=M_0$. On the
right-hand side of the figure, curves are marked from top to bottom
by $t = 0,~ 1.00\sqrt{\sigma},~ 2.00\sqrt{\sigma},~
3.00\sqrt{\sigma},$ and $4.00\sqrt{\sigma}$. The figure shows the
possibility of having an extremal configuration with one degenerate
event horizon. }}
 \label{fig:2}
\end{figure}

\begin{figure}[htp]
\begin{center}
\includegraphics{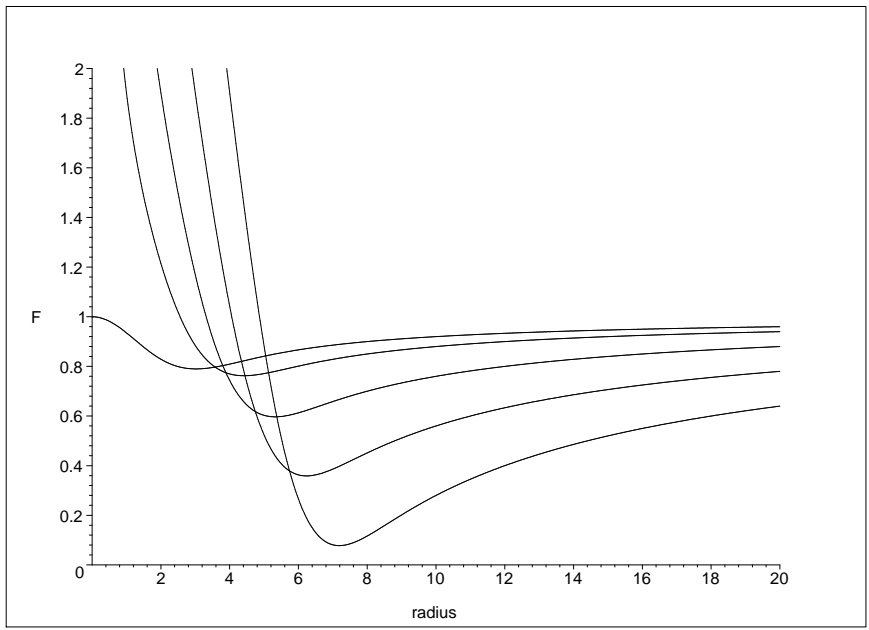}
\end{center}
\vspace{7.5 cm}
 \caption{\scriptsize {The temporal component of the
metric versus $\frac{r}{\sqrt{\sigma}}$ for different values of
$\frac{t}{\sqrt{\sigma}}$ with a sufficiently small amount of
initial mass ($M_I<M_0$), e.g. $M_I=0.40\sqrt{\sigma}$. On the
right-hand side of the figure, curves are marked from top to bottom
by $t = 0,~ 1.00\sqrt{\sigma},~ 2.00\sqrt{\sigma},~
3.00\sqrt{\sigma},$ and $4.00\sqrt{\sigma}$. The figure does not
show any event horizon when the initial mass of the black hole is
smaller than the minimal nonzero mass.  }}
 \label{fig:3}
\end{figure}

To compute the mass function, we consider the situation of nonstatic
in which the analytic mass solution is time dependent, $M=M(t,r)$.
Using Eq.~(\ref{mat:4}), the mass function is obtained by the
constraint $T_{r}^{\,\,\,\,r}=T_{\theta}^{\,\,\,\,\theta}$, which
gives
\begin{equation}
\label{mat:10}M=C\,e^{[\frac{t^2}{4\sigma}+\frac{t(r-t)}{2\sigma}]}.
\end{equation}
If we choose $C=M_I$ (initial black hole mass) to have physical
meaningful solutions, then plugging the above $M$ into the relation
(\ref{mat:5}), we find the line element:
\begin{equation}
\label{mat:11}ds^2=-F(t,r)dt^2+F^{-1}(t,r)dr^2+r^2d\Omega^2,
\end{equation}
with
\begin{equation}
\label{mat:12}F(t,r)=1-\frac{2M_\sigma(t,r)}{r},
\end{equation}
where the Gaussian-smeared mass distribution immediately reads
\begin{equation}
\label{mat:13}M_\sigma(t,r)=M_I\left(
{\cal{E}}\left(\frac{r-t}{2\sqrt{\sigma}}\right)\left(1+\frac{t^2}{2\sigma}\right)
-\frac{r}{\sqrt{\pi\sigma}}e^{-\frac{(r-t)^2}{4\sigma}}\left(1+\frac{t}{r}\right)\right).
\end{equation}
${\cal{E}}(x)$ shows the {\it Gauss error function} defined as $
{\cal{E}}(x)\equiv \frac{2}{\sqrt{\pi}}\int_{0}^{x}e^{-p^2}dp$. To
find the $F(t,r)$ we have set the value of integration constant to
zero. In the limit $\frac{t}{\sqrt{\sigma}}\gg1$ and also
$\frac{r}{\sqrt{\sigma}}\gg1$, the expression given in
Eq.~(\ref{mat:12}) satisfies Eq.~(\ref{mat:7}) with a good
approximation. Depending on the different values of initial mass
$M_I$, and upon a numerical solution, the metric displays three
possible causal structures: (1) It is possible to have two distinct
horizons when the initial mass of the black hole is larger than
minimal nonzero mass $M_0$, i.e. $M_I>M_0$ (see Fig.~\ref{fig:1}).
(2) It is possible to have one degenerate horizon (extremal black
hole), for $M_I=M_0$ (see Fig.~\ref{fig:2}). (3) It is impossible to
have a horizon at all (for $M_I<M_0$), and this possibility is shown
in Fig.~\ref{fig:3}.

As Fig.~\ref{fig:1} shows, for a sufficiently large and fixed
$\frac{M_I}{\sqrt{\sigma}}$, e.g. $M_I=3.00\sqrt{\sigma}$, the
distance between the horizons will increase as time progresses. The
appearance of a naked singularity at $r = 0$ in a nonstatic case is
natural which has not been supported by the cosmic censorship
conjecture \cite{pen}. One of the main arguments in support of the
censorship conjecture is the stability of black holes concerning
small perturbations. Indeed there are various effects that point out
that at least in its simplest form the censorship conjecture is
dubious. Let us consider the possible blueshift unstableness at the
inner horizon as an example which demonstrates the feasible
formation of naked singularities in conditions which might be
considered as physically sensible. Since an observer passing through
the inner horizon would encounter an infinite blueshift of any
entering emission, as he comes near the horizon, he surveys the
total history of the outward region in a determinate interval of his
individual proper time. So it would be possible for any small
perturbation to upset the horizon and emerge as a naked singularity.
In Ref.~\cite{nam} it was illustrated that the inner (Cauchy)
horizon of some of the black holes (e.g. Reissner-Nordstr\"{o}m
black holes) is unsteady and their solutions are unsuccessful to be
globally hyperbolic. If a spacetime fails to be globally hyperbolic,
the weak censorship conjecture is destroyed \cite{tip}. Then it must
include a naked singularity, i.e., there exists a future directed
causal curve that arrives a distant observer, and in the past it
ends at the singularity.

On the other hand when \,$r$ \,is immoderately small, in the region
where noncommutativity effects accurately commence to be perceived,
the detailed nature of the sharpened mass distribution is not
practically being scrutinized. Recently \cite{meh1}, we have
reported some results about extraordinary thermodynamical behavior
for Planck-scale black hole evaporation, i.e., when $M_I$ is less
than $M_0$, where there is the principal reactiveness to
noncommutativity effects and the detailed form of the matter
distribution. In this area, some unusual thermodynamical features,
e.g., negative entropy, negative temperature, and abnormal heat
capacity appeared. There are also the predominant differences
between the Gaussian, Lorentzian, or some other forms of the smeared
mass distribution at this extreme regime. In other words, the bases
of these theories probably become as a result of the fractal nature
of spacetime at very short distances. Theories such as $E$ infinity
\cite{nas} and scale relativity \cite{not} which are on the basis of
the fractal structure of spacetime at very short distances may
provide a suitable framework to handle thermodynamics of these very
short distance systems. Therefore, we really should not have
credence to the details of our modeling when
\,$\frac{r}{\sqrt{\sigma}}\ll1$ and only apply the Gaussian-smeared
mass distribution in our calculations just on the condition that
$M_I\geq M_0$.

The plot presented in Fig.~\ref{fig:2} shows, for several values of
minimal nonzero mass $M_0$, the possibility of having an extremal
configuration with one degenerate event horizon as time progresses.
For more details, the numerical results for the remnant size of the
black hole for different values of $\frac{t}{\sqrt{\sigma}}$ are
presented in Table~\ref{tab:1}. According to Table~\ref{tab:1}, as
time moves forward the minimal nonzero mass decreases but the
minimal nonzero horizon radius increases which means that in the
limit $\frac{t}{\sqrt{\sigma}}\gg1$, the micro black hole can
evaporate completely, i.e. $M_0\rightarrow 0$. Therefore, the idea
of a stable black hole remnant as a candidate to conserve
information has failed. Note that, currently there are some
proposals about what happens to the information that falls into a
black hole. One of the main proposals is that the black hole never
disappears completely, and the information is not lost, but would be
stored in a Planck size stable remnant (for reviews on resolving the
so-called information loss problem, see \cite{pre}).

In fact, when one considers the time-varying mass, based on our
model and preferred calculations, total evaporation of the black
hole is possible in principle. This is in agreement with the
original Bekenstein-Hawking approach \cite{haw,bek} and also our
approach by
using the time-varying speed of light model \cite{noz}.\\
\begin{table}
\caption{The minimal nonzero mass of the black hole (remnant mass,
$\frac{M_0}{\sqrt{\sigma}}$) and also the minimal nonzero horizon
radius, $\frac{r_0}{\sqrt{\sigma}}$, for different values of
$\frac{t}{\sqrt{\sigma}}$. In the long-time limit, i.e.
$\frac{t}{\sqrt{\sigma}}\gg1$, there is no black hole remnant.}
\begin{center}
\begin{tabular}{|c|c|c|}
\hline
\multicolumn{3}{|c|}{Extremal black hole} \\
\hline Time & Minimal nonzero mass & Minimal nonzero horizon radius  \\
\hline$t=0$ & $M_0\approx1.90\sqrt{\sigma}$ & $r_0\approx3.02\sqrt{\sigma}$\\
\hline$t=1.00\sqrt{\sigma}$ & $M_0\approx1.68\sqrt{\sigma}$ & $r_0\approx4.49\sqrt{\sigma}$ \\
\hline$t=2.00\sqrt{\sigma}$ & $M_0\approx0.99\sqrt{\sigma}$ & $r_0\approx5.34\sqrt{\sigma}$ \\
\hline$t=3.00\sqrt{\sigma}$ & $M_0\approx0.62\sqrt{\sigma}$ & $r_0\approx6.14\sqrt{\sigma}$\\
\hline$t=4.00\sqrt{\sigma}$ & $M_0\approx0.43\sqrt{\sigma}$ & $r_0\approx7.18\sqrt{\sigma}$\\
\hline$t=5.00\sqrt{\sigma}$ & $M_0\approx0.32\sqrt{\sigma}$ & $r_0\approx8.32\sqrt{\sigma}$\\
\hline$t=10.00\sqrt{\sigma}$ & $M_0\approx0.13\sqrt{\sigma}$ & $r_0\approx13.27\sqrt{\sigma}$\\
\hline$t=100.00\sqrt{\sigma}$ & $M_0\approx0.01\sqrt{\sigma}$ & $r_0\approx105.05\sqrt{\sigma}$\\
\hline$t\rightarrow\infty$ & $M_0\rightarrow0$ & $r_0\rightarrow\infty$\\
\hline
\end{tabular}
\end{center}
\label{tab:1}
\end{table}
For a sufficiently small and fixed $\frac{M_I}{\sqrt{\sigma}}$, e.g.
$M_I=0.40\sqrt{\sigma}$, there is no event horizon when the initial
mass of the black hole is smaller than the minimal nonzero mass
which is shown in Fig.~\ref{fig:3}.

\begin{figure}[htp]
\begin{center}
\includegraphics{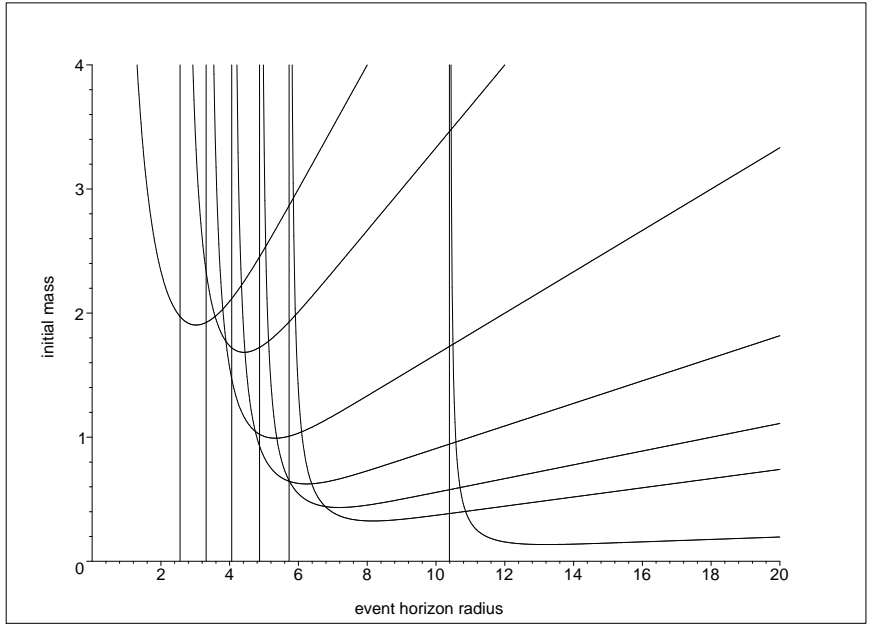}
\end{center}
\vspace{8 cm} \caption{\scriptsize {The initial mass,
$\frac{M_I}{\sqrt{\sigma}}$, versus the event horizon radius,
$\frac{r_H}{\sqrt{\sigma}}$, for different values of
$\frac{t}{\sqrt{\sigma}}$. On the right-hand side of the figure,
curves are marked from top to bottom by $t = 0,~ 1.00\sqrt{\sigma},~
2.00\sqrt{\sigma},~ 3.00\sqrt{\sigma},~ 4.00\sqrt{\sigma},~
5.00\sqrt{\sigma},$ and $10.00\sqrt{\sigma}$. As can be seen from
the figure, the results are similar to Table~\ref{tab:1}. }}
\label{fig:4}
\end{figure}

Finally, the solution (\ref{mat:12}) can be substituted in
(\ref{mat:6}) to obtain a solution for $T_{t}^{\,\,\,\,r}$ as the
following expression:
\begin{equation}
\label{mat:14}T_{t}^{\,\,\,\,r}=\frac{-\sqrt{\sigma}e^{-\frac{(r-t)^2}{4\sigma}}(r^2+4\sigma)+2\sqrt{\pi}\sigma
t{\cal{E}}\left(\frac{r-t}{2\sqrt{\sigma}}\right)}{8\sqrt{\pi}M_I\left[2\sqrt{\sigma}e^{-\frac{(r-t)^2}{4\sigma}}(r+t)
-\sqrt{\pi}{\cal{E}}\left(\frac{r-t}{2\sqrt{\sigma}}\right)(t^2+2\sigma)+\sqrt{\pi}\sigma
r\right]^2}.
\end{equation}
For the time-independent case, $t=0$, one recovers the
noncommutative-Schwarzschild case \cite{nic}, i.e.,
\begin{equation}
\label{mat:15}F(r)=1-\frac{2M_I}{r}
{\cal{E}}\left(\frac{r}{2\sqrt{\sigma}}\right)+\frac{2M_I}{\sqrt{\pi\sigma}}e^{-\frac{r^2}{4\sigma}},
\end{equation}
and $T_{t}^{\,\,\,\,r}=0$. In the commutative limit concerning the
above equation, $\sigma\rightarrow0$, the Gauss error function tends
to 1 and the other term will exponentially be reduced to zero. Thus
one retrieves the conventional result
\begin{equation}
\label{mat:16}F(r)=1-\frac{2M_I}{r}.
\end{equation}
Therefore, the modified Vaidya solution is reduced to the ordinary
Schwarzschild solution. It is clear that the line element
(\ref{mat:11}) has a coordinate singularity at the event horizon as
\begin{equation}
\label{mat:17}r_H = 2M_\sigma(t,r_H).
\end{equation}
The analytical solution of Eq. (\ref{mat:17}) for $r_H$ in a closed
form is impossible, but it is possible to solve (\ref{mat:17}) to
find $M_I$, which provides the initial mass as a function of the
horizon radius $r_H$. This leads to
\begin{equation}
\label{mat:18}M_I=\sqrt{\pi}\sigma^{\frac{3}{2}}r_H\left[\sqrt{\pi\sigma}{\cal{E}}\left(\frac{r_H-t}{2\sqrt{\sigma}}\right)
\left(2\sigma+t^2\right)-2\sigma
e^{-\frac{(r_H-t)^2}{4\sigma}}\left(r_H+t\right)\right]^{-1}.
\end{equation}
The results of the numerical solution of the initial mass as a
function of the horizon radius are displayed in Fig.~\ref{fig:4}
which are comparable to Table~\ref{tab:1}.
As expected, from the initial mass equation (see Fig.~\ref{fig:4}),
one acquires that noncommutativity indicates a minimal nonzero mass
in order to have an event horizon. So, in the noncommutative case,
for $M_I < M_0$ there is no event horizon (see Fig.~\ref{fig:2} and
also Table~\ref{tab:1}).\\

\section{\label{sec:3}Parikh-Wilczek Tunneling}
\begin{figure}[htp]
\begin{center}
\includegraphics{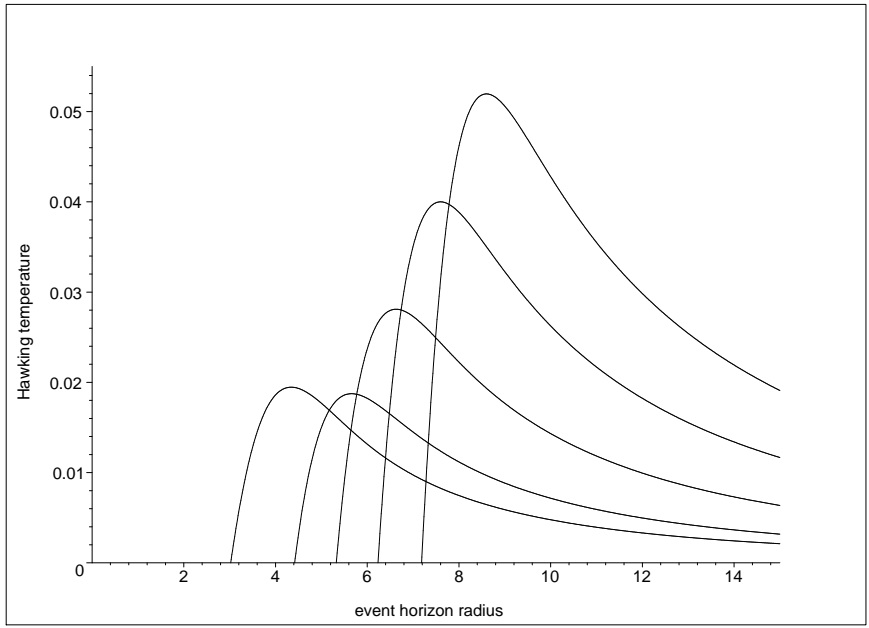}
\end{center}
\vspace{8 cm}
 \caption{\scriptsize {The Hawking temperature, $T_H\sqrt{\sigma}$, as a function of horizon radius (the outer horizon radius),
$\frac{r_H}{\sqrt{\sigma}}$. We have set $M_I=3.00\sqrt{\sigma}$.
The existence of a minimal nonzero mass and disappearance of
divergence are clear. On the right-hand side of the figure, curves
are marked from bottom to top by $t =
 0,~ 1.00\sqrt{\sigma},~ 2.00\sqrt{\sigma},~ 3.00\sqrt{\sigma},$ and $4.00\sqrt{\sigma}$.}}
 \label{fig:5}
\end{figure}

The radiating behavior associated with this noncommutative black
hole solution can now be investigated by the quantum tunneling
process suggested by Parikh and Wilczek \cite{par1}. To portray the
quantum tunneling method where a particle moves in dynamical
geometry and passes through the horizon without singularity on the
path, we should utilize a coordinate system that is not singular at
the horizon. Painlev\'{e} coordinates \cite{pai} which are used to
eliminate coordinate singularity are especially convenient choices
in this analysis. Under the Painlev\'{e} time coordinate
transformation,
\begin{equation}
\label{mat:19} dt\rightarrow dt-\frac{\sqrt{1-F(t,r)}}{F(t,r)}dr,
\end{equation}
the noncommutative Painlev\'{e} metric now takes the following form:
$$ds^2=-F(t,r)dt^2+2\sqrt{1-F(t,r)}dtdr+dr^2+r^2d\Omega^2$$
\begin{equation}
\label{mat:20}=-\left(1-\frac{2M_\sigma(t,r)}{r}\right)dt^2+2\sqrt{\frac{2M_\sigma(t,r)}{r}}dtdr+dr^2+r^2d\Omega^2.
\end{equation}
The metric is now stationary, and there is no coordinate singularity
at the horizon. The outgoing motion of the massless particles (the
outgoing radial null geodesics, $ds^2=d\Omega^2=0$) takes the form
\begin{equation}
\label{mat:21} \frac{dr}{dt}=1-\sqrt{1-F(t,r)}.
\end{equation}
Since we just need an approximation value of $F(t,r)$ for short
distances in the vicinity of the horizon, we can expand the
coefficient $F(t,r)$ by using the Taylor series at a fixed time and
just to first order. So, we have
\begin{equation}
\label{mat:22} F(t,r)\Big|_t=
F'(t,r_H)\Big|_t\,(r-r_H)+O\left((r-r_H)^2\right)\Big|_t.
\end{equation}
By this approximation at the neighborhood of the black hole horizon,
the equation of radial null geodesic can be obtained by
\begin{equation}
\label{mat:23}
\frac{dr}{dt}\simeq\frac{1}{2}F'(t,r_H)\,(r-r_H)\simeq\kappa(M_I)\,(r-r_H),
\end{equation}
where $\kappa(M_I)\simeq\frac{1}{2}F'(t,r_H)$ is the surface gravity
for the metric (\ref{mat:20}) at the horizon. Now we are ready to
consider the Hawking temperature of such a black hole (see
\cite{sia} for a more detailed discussion of the semiclassical
methods to derive the Hawking temperature in the Vaidya black hole).
From the expression
$T_H=\frac{\kappa}{2\pi}=\frac{1}{4\pi}F'(t,r_H)\Big|_t$, the
noncommutative Hawking temperature including the time-dependent part
is given by
\begin{equation}
\label{mat:24}T_H=M_I\left[\frac{{\cal{E}}\left(\frac{r_H-t}{2\sqrt{\sigma}}\right)}{2\pi
r_H^2}
\left(1+\frac{t^2}{2\sigma}\right)-\frac{e^{-\frac{(r_H-t)^2}{4\sigma}}}{4(\pi\sigma)^{\frac{3}{2}}}\left(r_H+\frac{2\sigma}{r_H}
+\frac{2\sigma t}{r_H^2} \right)\right].
\end{equation}
For the time-independent case, $t=0$, one retrieves the Hawking
temperature for the noncommutative Schwarzschild black hole that is
consistent with the Ref.~\cite{nic}. In the limit of $\sigma$ going
to zero, we get the classical Hawking temperature,
$T_H=\frac{1}{8\pi M_I}$. The numerical computation of the
noncommutative Hawking temperature as a function of horizon radius
(the outer horizon radius) is depicted in Fig.~\ref{fig:5}. As can
be seen from Fig.~\ref{fig:5} the black hole at the ultimate phase
of evaporation ceases to radiate, its temperature reaches zero and
the existence of a minimal nonzero mass is clear. In this modified
version, there is no divergence at the final stage of the black hole
evaporation because the temperature reaches a maximum definite value
before cooling down to absolute zero, at the minimal nonzero value
of the outer horizon radius $r_0$, that the black hole shrinks to
(see Table~\ref{tab:1}).

Let us come back to the tunneling procedure. In accordance with the
original work proposed by Parikh and Wilczek \cite{par1}, the WKB
approximation is valid at the neighborhood of the horizon. Then, the
emission rate for the classically forbidden region as a function of
the imaginary part of the action for a particle in a tunneling
process is given by {\footnote {We should stress that there is
another point to using relation (\ref{mat:25}). There is a problem
here known as the "factor $2$ problem" \cite{akh}. Recently, some
authors (see \cite{cho,pil} and references therein) have declared
that the relation (\ref{mat:25}) is not invariant under canonical
transformations but the same formula with a factor of $1/2$ in the
exponent is canonically invariant. This method leads to a
temperature which is higher than the Hawking temperature by a factor
of $2$. In Ref.~\cite{akh2}, a resolution to this problem was given
in terms of an overlooked temporal contribution to the tunneling
amplitude. When one includes this temporal contribution one gets
exactly the correct temperature and exactly when one uses the
canonically invariant tunneling amplitude.}}
\begin{equation}
\label{mat:25}\Gamma\sim e^{-2\textmd{Im}\,\textit{I}}.
\end{equation}
Now, we consider a spherical positive energy shell containing the
components of massless particles each of which journeys on a radial
null geodesic like an $s$-wave outgoing particle which crosses the
horizon in the outward direction from $r_{in}$ to $r_{out}$. Hence,
the imaginary part of the action takes the following form
\begin{equation}
\label{mat:26}\textmd{Im}\,
I=\textmd{Im}\int_{r_{in}}^{r_{out}}p_rdr=\textmd{Im}\int_{r_{in}}^{r_{out}}\int_0^{p_r}dp'_rdr.
\end{equation}
Utilizing Hamilton's equation of motion
$\frac{dr}{dt}=\frac{dH}{dp_r}|_r$, the integral variable is changed
from momentum to energy. So, we have
\begin{equation}
\label{mat:27}\textmd{Im}\,
I=\textmd{Im}\int_{r_{in}}^{r_{out}}\int_{0}^{H}\frac{dH'}{\frac{dr}{dt}}dr.
\end{equation}
If we consider the particle's self-gravitation effect, according to
the original work by Kraus and Wilczek \cite{kra}, then
Eq.~(\ref{mat:23}) should be modified. We retain the total
Arnowitt-Deser-Misner mass ($M_I$) of the spacetime fixed, and allow
the hole mass to fluctuate because we take into consideration the
response of the background geometry which corresponds to an emitted
quantum of energy $E$ at a fixed time or a stationary phase.
Therefore we should replace $M_I$ by $M_I - E$ in Eq.~(\ref{mat:23})
and then (\ref{mat:27}). The imaginary action (\ref{mat:27}) now
becomes
\begin{equation}
\label{mat:28}\textmd{Im}\,
I=\textmd{Im}\int_{r_{in}}^{r_{out}}\int_{M_I}^{M_I-E}\frac{d(M_I-E')}{\kappa(M_I-E')\,(r-r_H)}dr
=-\textmd{Im}\int_{r_{in}}^{r_{out}}\int_{0}^{E}\frac{dE'}{\kappa(M_I-E')\,(r-r_H)}dr.
\end{equation}
The $r$ integral can be performed first by a contour integration for
the lower half $E'$ plane due to the escape from the pole at the
horizon. In this way, we acquire
\begin{equation}
\label{mat:29}\textmd{Im}\,
I=-\textmd{Im}\int_{0}^{E}\frac{dE'}{\kappa(M_I-E')}\int_{r_{in}}^{r_{out}}\frac{dr}{r-r_H}=
\pi\int_{0}^{E}\frac{dE'}{\kappa(M_I-E')},
\end{equation}
on the condition that $r_{in} > r_{out}$. Using the first low of
black hole thermodynamics, $dM=\frac{\kappa}{2\pi}dS$, the
expression \,$\textmd{Im}\,I$ given by \cite{kes}
\begin{equation}
\label{mat:30}\textmd{Im}\,I=-\frac{1}{2}\int_{S_{NC}(M_I)}^{S_{NC}(M_I-E)}dS=-\frac{1}{2}\Delta
S_{NC},
\end{equation}
where $S_{NC}$ is the noncommutative black hole entropy. The
tunneling amplitude in the high energy depends on the final and
initial number of microstates available for the system (see also
\cite{arz,ban,ban2,mas}). Thus, we have
\begin{equation}
\label{mat:31}\Gamma\sim\frac{e^{S_{final}}}{e^{S_{initial}}}=
e^{\Delta S_{NC}}=e^{S_{NC}(M_I-E)-S_{NC}(M_I)}.
\end{equation}
From this viewpoint the emission rate is proportional to the
difference in black hole entropies before and after emission which
means that the emission spectrum cannot be accurately thermal at
higher energies.

We should note that the tunneling amplitude can also be obtained by
writing out the explicit metric in the tunneling calculation. To
find the analytic form of the difference in black hole entropies
before and after emission and then compute the expression $\Gamma$,
we evaluate the integral (\ref{mat:27}) by writing the explicit form
for the radial null geodesic, Eq.~(\ref{mat:21}), which incorporates
the backreaction effects. Therefore after performing the $r$
integration in Eq.~(\ref{mat:27}) by deforming the contour
{\footnote{Note that since there is no analytical solution for $r_H$
versus $M_I$, then one can approximately calculate the
noncommutative horizon radius versus the initial mass by setting
$r_H=2M_I$ into the smeared mass distribution $M_\sigma(t,r_H)$.}},
we find
\begin{equation}
\label{mat:32}\textmd{Im}\, I=\textmd{Im}\int_{0}^{E}4\pi i
M_\sigma(t, M_I-E')dE',
\end{equation}
where
$$M_\sigma\left(t,M_I-E\right)=(M_I-E)\Bigg[
{\cal{E}}\left(\frac{2(M_I-E)-t}{2\sqrt{\sigma}}\right)\left(1+\frac{t^2}{2\sigma}\right)$$
\begin{equation}
\label{mat:33}
-\frac{2(M_I-E)}{\sqrt{\pi\sigma}}e^{-\frac{\left(2(M_I-E)-t\right)^2}{4\sigma}}\left(1+\frac{t}{2(M_I-E)}\right)\Bigg],
\end{equation}
so we can find the noncommutative-corrected tunneling amplitude as
follows:
$$\Gamma\sim \exp(\Delta S_{NC})=\exp\Bigg({\cal{E}}\left(\frac{2(M_I-E)-t}{2\sqrt{\sigma}}\right)\Bigg[4\pi
(M_I-E)^2\bigg(1+\frac{t^2}{2\sigma}\bigg)$$$$-6\pi\left(\sigma+t^2\right)-\frac{\pi
t^4}{2\sigma}\Bigg]
+e^{-\frac{(2(M_I-E)-t)^2}{4\sigma}}\Bigg[2\sqrt{\pi\sigma}\left(6+5t\right)
+\sqrt{\frac{\pi}{\sigma}}t^2\left(2(M_I-E)+t\right)\Bigg]$$$$-{\cal{E}}\left(\frac{2M_I-t}{2\sqrt{\sigma}}\right)\left[4\pi
M^2_I\bigg(1+\frac{t^2}{2\sigma}\bigg)-6\pi\left(\sigma+t^2\right)-\frac{\pi
t^4}{2\sigma}\right]$$
\begin{equation}
\label{mat:34}-e^{-\frac{(2M_I-t)^2}{4\sigma}}\Bigg[2\sqrt{\pi\sigma}\left(6+5t\right)
+\sqrt{\frac{\pi}{\sigma}}t^2\left(2M_I+t\right)\Bigg]\Bigg).
\end{equation}
In this situation, we would like to test our result approximately.
It is adequate to acquire the analytic form of the noncommutative
entropy $S_{NC}$, and then compute the difference in black hole
entropies before and after emission, $\Delta
S_{NC}=S_{NC}(M_I-E)-S_{NC}(M_I)$, to compare between the first law
of black hole thermodynamics and the tunneling approaches. For this
purpose, we should note that our calculations to find
Eq.~(\ref{mat:24}) (noncommutative Hawking temperature) are accurate
and no approximation has been performed. But there is no analytical
solution for entropy from the first law of classical black hole
thermodynamics $dM=T_HdS$, even if we set $r_H=2M_I$ in
Eq.~(\ref{mat:24}). Now, we want to calculate the Hawking
temperature in an approximate way to find the analytical form of the
entropy as follows:
\begin{equation}
\label{mat:35}T_H=\frac{1}{4\pi r_H},
\end{equation}
where $r_H$ is given by
\begin{equation}
\label{mat:36}r_H = 2M_I\left(
{\cal{E}}\left(\frac{2M_I-t}{2\sqrt{\sigma}}\right)\left(1+\frac{t^2}{2\sigma}\right)
-\frac{2M_I}{\sqrt{\pi\sigma}}e^{-\frac{(2M_I-t)^2}{4\sigma}}\left(1+\frac{t}{2M_I}\right)\right).
\end{equation}
Finally, the entropy of the black hole can be achieved as the
analytical form by using the first low of classical black hole
thermodynamics,
$$S_{NC}=\int\frac{dM_I}{T_H}={\cal{E}}\left(\frac{2M_I-t}{2\sqrt{\sigma}}\right)\left[4\pi
M^2_I\bigg(1+\frac{t^2}{2\sigma}\bigg)-6\pi\left(\sigma+t^2\right)-\frac{\pi
t^4}{2\sigma}\right]$$
\begin{equation}
\label{mat:37}+e^{-\frac{(2M_I-t)^2}{4\sigma}}\Bigg[2\sqrt{\pi\sigma}\left(6+5t\right)
+\sqrt{\frac{\pi}{\sigma}}t^2\left(2M_I+t\right)\Bigg].
\end{equation}
It is clear that the two approaches mentioned above are accurately
coincided, and it can be easily checked by computing $\Delta
S_{NC}$; however our approach has been approximated. We should
stress that the total of our results in Ref.~\cite{meh2} are
recovered by setting $t=0$ into the above corresponding equations.

The question which arises here is the possible dependences between
different modes of radiation during the evaporation and then the
time evolution of these possible correlations which needs further
investigation and probably sheds more light on the information loss
problem \cite{pre}. This problem is currently under investigation.\\

\section{\label{sec:4}Summary}
In summary, we have analyzed a solution of the Einstein equations
with a noncommutative distribution of mass/energy which is
spherically symmetric, time-dependent and localized near the origin
of the spacetime, namely, the noncommutative-inspired Vaidya
solution. In this setup, the proposal of stable black hole remnant
as a candidate to store information has failed because the black
hole evaporates completely in the long time limit. Finally, by using
the semiclassical method, we have derived the Hawking temperature
and the emission rate via a tunneling process which includes the
corrections due to noncommutativity. The entropy for such a black
hole is approximately computed in a closed form. These corrections
would be significant at the level of semiclassical quantum gravity,
specifically once the black hole mass becomes close
to the Planck mass.\\

\end{document}